\documentclass[aps,showpacs,preprintnumbers]{revtex4}
\usepackage{amssymb}
\usepackage{bm}
\usepackage{fancyhdr}
\usepackage[latin1]{inputenc}
\usepackage{amsmath}
\usepackage{empheq,amsmath}
\usepackage[brazil]{babel}

\def\M{\mathbb{M}}

\def\R{\mathbb{R}}

\def\M2{\R^{2 \times 2}}

\def\||{\parallel}

\def\qed{\vbox{\hbox to \textwidth{\hbox{} \rightline{\slshape q.e.d.} \hfil}}}
\input{tcilatex}
\begin{document}

\title{Angular Tunneling Effect}
\author{Cássio Lima$^{1}$, Jorge Henrique Sales$^{1}$, A. T. Suzuki$^{2}$.}
\affiliation{$^{1}$Universidade Estadual de Santa Cruz, Departamento de Ci\^encias Exatas
e Tecnol\'ogicas, 45662-000 - Ilh\'eus, BA, Brasil}
\affiliation{$^{2}$Instituto de Física Te\'orica, Universidade Estadual Paulista\\
Rua Dr. Bento Teobaldo Ferraz, 271 -- 01140-070 --- S\~ao Paulo, SP -
Brazil. }
\date{\today }

\begin{abstract}
We investigate the tunneling of an electron with momentum $p$ in the
direction of $V$ potential and under an angle $\theta $ to the normal
potential. Using the boundary conditions, the conditions of continuity and
Snell's law, we obtain tunneling for various angles.
\end{abstract}

\maketitle

\section{Introduction}

The alpha-decay process was interpreted in the early 1920s in terms of
tunnelling through a quantum mechanical potential barrier \cite[2]{1}.
Nuclear deformation e ects have been studied in 1950s by Bohr et al \cite{3}
and Froman \cite{4}. Recently, two theoretical extreme approaches have been
developed to describe the alpha decay: the cluster- and ssion-like theories 
\cite{5}. A lot of new experimental and theoretical investigation on alpha
decay half-life has been developed during the last three years or so \cite[%
7,8,10,11,12,13,14,15,16,17,18,19]{6}. In addition, half-life values for
spontaneous nuclear decay processes (proton emission, alpha decay, cluster
radioactivity, and cold ssion) have been presented very recently in the
framework of the Effective Liquid Drop Model \cite{22}.

\bigskip

Penetration property in a classically forbidden region allows the
understanding of several phenomena such as electron tunneling and alpha
decay, from the analysis of the behavior of a particle concerning a
potential barrier rectangular figure 1, where it can be transmitted and
reflected.

A potential barrier is defined by

\begin{equation}
V(x)=\left\{ 
\begin{array}{ccc}
V & \mbox{ se } & 0\leq x\leq a \\ 
0 & \mbox{ se } & x<0\ \mbox{ ou }\ x>a%
\end{array}%
\right. \ .  \label{1}
\end{equation}

According to classical physics, a particle of energy $E$ less than the
height $V$ of a barrier could not penetrate -- the region inside the barrier
is classically forbidden. But the wavefunction associated with a free
particle must be continuous at the barrier and will show an exponential
decay inside the barrier. The wavefunction must also be continuous on the
far side of the barrier, so there is a finite probability that the particle
will tunnel through the barrier. The transmission coefficient T is the
probability of a particle incident from the left (region 1) to be tunneling
through the barrier (region 2) and continue to travel to the right (region
3) is given by

\bigskip 
\begin{equation}
\Im \simeq 16\displaystyle\frac{E}{V}\left( 1-\displaystyle\frac{E}{V}%
\right) e^{-2k_{2}a}.  \label{eq25}
\end{equation}

\bigskip

Equation (\ref{eq25}) shows us that a particule with mass $m$ and energy $%
E<V $, that approaches a potencial barrier $V$ and $a$ wide has a
probability $\Im \neq 0$ of penetrated the barrier and appear on the other
side. This phenomena is known as tunneling.

\section{\protect\bigskip Angular Tunneling}

\bigskip

The model presented in this section serves for scattering, collimated and
nearly monoenergetic beam of particles, dispersions that wave packets are
very small, and the reflection and transmission coefficients can be
determined from the components of the propagation of monochromatic plane
wave eigenfunctions the Hamiltonian of the free particle, with free energy $%
E $. Thus, if the particle of mass $m$ approaches the barrier around the
region $1($Fig.1) with incidence angle $\theta _{1}$with normal line to the
barrier surface, the incident and reflected state is represented by the wave 
$\psi _{1}(r)$

\begin{equation}
\psi _{1}(r)=Ae^{i\overrightarrow{k_{1}}\cdot \overrightarrow{r}}+Be^{-i%
\overrightarrow{k_{1}}\cdot \overrightarrow{r}}  \label{2}
\end{equation}%
where $\overrightarrow{k_{1}}\cdot \overrightarrow{r}=k_{1}r\mathrm{\;%
\mathrm{\;cos}\;}\theta _{1}$, being the vector $\overrightarrow{r}$ any
position in the plane that separates two regions $1$ and $2$ Fig.\ref{fig2}
e $\overrightarrow{k}_{1}=\frac{\overrightarrow{p}_{1}}{\hbar }$. The second
installment of the wave function is the state associated with the reflection
of the particle through the barrier.

The wave function which crosses the barrier is given by%
\begin{equation}
\psi _{2}(r)=Ce^{i\overrightarrow{k}\cdot \overrightarrow{r}}+De^{-i%
\overrightarrow{k}\cdot \overrightarrow{r}},\qquad \ \overrightarrow{k}\cdot 
\overrightarrow{r}=kr\mathrm{\mathrm{\;cos}\;}\theta _{2}  \label{3}
\end{equation}%
and the transmitted wave function, passing the region 2 to region 3 is given
by

\begin{equation}
\psi _{3}(r)=Ee^{i\overrightarrow{k_{1}}\cdot \overrightarrow{r}},\ \ \ 
\overrightarrow{k_{1}}\cdot \overrightarrow{r}=k_{1}r\mathrm{\mathrm{\; cos}%
\; }\theta _{3}  \label{4}
\end{equation}

\bigskip Assuming that the absolute refractive index for region 1 is $n_{1}=%
\frac{c}{v_{1}}$\ and it is equal to the region 3, i.e., $n_{1}=n_{3}$, this
implies by Snell's refraction equation $\theta _{3}=\theta _{1}.$So, for
region 3 the emerging wave function from barrier 3 $\psi _{3}(r)$ depends on 
$\theta _{1}$ incidence angle.

Thus, for region 3, the emerging wave function from barrier $\psi _{3}(r)$ \
depends on $\theta _{1}$ incidence angle.

\bigskip

First of all lets compute for the case of the step potential, $E>V$, and we
need to treat the problem using the boundary conditions defined at $r=0$ and 
$r=a$. The current probability must remain continuous at the origin, despite
the discontinuity of potential. This implies that the wave function and its
first derivative with respect to $r$ should be continuous in the case $r=0$

\begin{equation}
\Psi _{1}(0)=\Psi _{2}(0)\Longrightarrow A+B=C+D  \label{s1}
\end{equation}

The first derivative of the wave functions are

\begin{eqnarray}
\frac{\partial \Psi _{1}}{\partial r} &=&ik_{1}r\mathrm{\mathrm{\; cos}\; }%
\theta _{1}\left( Ae^{i\overrightarrow{k_{1}}\cdot \overrightarrow{r}}+Be^{-i%
\overrightarrow{k_{1}}\cdot \overrightarrow{r}}\right)  \label{5} \\
\frac{\partial \Psi _{2}}{\partial r} &=&ikr\mathrm{\mathrm{\; cos}\; }%
\theta _{2}\left( Ce^{i\overrightarrow{k}\cdot \overrightarrow{r}}-De^{-i%
\overrightarrow{k}\cdot \overrightarrow{r}}\right)  \notag \\
\frac{\partial \Psi _{3}}{\partial r} &=&ik_{1}r\mathrm{\mathrm{\; cos}\; }%
\theta _{1}\left( Ee^{i\overrightarrow{k_{1}}\cdot \overrightarrow{r}}\right)
\notag
\end{eqnarray}

for $r=0$, we have

\bigskip

\begin{eqnarray*}
ik_{1}\mathrm{\mathrm{\;cos}\;}\theta _{1}(A-B) &=&ik_{2}\mathrm{\mathrm{%
\;cos}\;}\theta _{2}(C-D) \\
A-B &=&\frac{1}{n}\frac{\mathrm{\;\mathrm{\;cos}\;}\theta _{2}}{\mathrm{\;%
\mathrm{\;cos}\;}\theta _{1}}(C-D)\text{ }
\end{eqnarray*}%
where

\begin{equation*}
n=\frac{k_{1}}{k_{2}}
\end{equation*}

\bigskip In the case $r=a$%
\begin{equation*}
Z_{1}C+\frac{1}{Z_{2}}D=Z_{1}E\text{ }
\end{equation*}

\bigskip For the first derivative at $r=a$

\bigskip

\begin{eqnarray*}
ik_{2}\mathrm{\mathrm{\;cos}\;}\theta _{2}CZ_{2}-\frac{ik_{2}\mathrm{\mathrm{%
\;cos}\;}\theta _{2}}{Z_{2}}D &=&iEZ_{1}k_{1}\mathrm{\mathrm{\;cos}\;}\theta
_{1} \\
CZ_{2}-\frac{1}{Z_{2}}D &=&Z_{1}En\frac{\mathrm{\;\mathrm{\;cos}\;}\theta
_{1}}{\mathrm{\;\mathrm{\;cos}\;}\theta _{2}}
\end{eqnarray*}%
and%
\begin{eqnarray*}
Z_{1} &=&e^{ik_{1}a\mathrm{\;\mathrm{\;cos}\;}\theta _{1}} \\
Z_{2} &=&e^{ik_{2}a\mathrm{\;\mathrm{\;cos}\;}\theta _{2}}
\end{eqnarray*}

Solving the equation systems, we have

\bigskip

\begin{equation*}
A=\frac{1}{4N}\frac{Z_{1}}{Z_{2}}\left( 1+N\right) ^{2}\left[ 1-\frac{\left(
N-1\right) ^{2}}{\left( N+1\right) ^{2}}\cdot Z_{2}^{2}\right] E
\end{equation*}%
where $N=n\alpha $ and $\alpha =\frac{\mathrm{\;cos}\;\theta _{1}}{\mathrm{%
\;cos}\;\theta _{2}}$. In the limit of $\theta _{1}=\theta _{2}=0$, $N$
tends to the refractive index $n$.

The calculation of the amplitude for the barrier is given by:

\begin{equation*}
T=\frac{E}{A}
\end{equation*}

\begin{equation*}
T=\frac{4NZ_{2}}{Z_{1}}\frac{1}{\left( N+1\right) ^{2}}\frac{1}{\left[
1-\left( \frac{N-1}{N+1}\right) ^{2}Z_{2}^{2}\right] }
\end{equation*}

\bigskip

or

\bigskip

\begin{equation}
T=\frac{4N}{\left( 1+N\right) ^{2}}\frac{e^{i(k_{2}-k_{1})a}}{\left[
1-\left( \frac{N-1}{N+1}\right) ^{2}\right] e^{2ik_{2}a}}  \label{step}
\end{equation}

In the case for tunneling: $0<E<V$

\bigskip

\begin{equation*}
V(x)=\left\{ 
\begin{array}{c}
0\text{, }x<0\longrightarrow \text{region 1} \\ 
V>0\text{, }0<x<a\rightarrow \text{ region 2} \\ 
0\text{, }x>a\rightarrow \text{ region 3}%
\end{array}%
\right\}
\end{equation*}

The above solutions are still valid, now we just take

\begin{equation*}
k_{2}=i\left\vert k_{2}\right\vert =\frac{i}{\hbar }\sqrt{2m(V-E)}\equiv iK,%
\text{ \ \ \ \ }k>0
\end{equation*}%
The wave equation for region $2$ is

\bigskip

\begin{equation*}
\psi _{2}(r)=Ce^{-kr}+De^{kr}\qquad \ (0\leq r\leq a)
\end{equation*}

\bigskip

\bigskip

but here, unlike the step, one can not exclude the exponentially increasing
because $r$ does not extend to -$\infty $ (boundary region) because we have
a limit to the potential that is up to$r=a$. Another important fact is that
the index of refraction becomes imaginary, i.e.

\bigskip

\begin{equation*}
n=\frac{k_{1}}{k_{2}}=i\frac{k_{1}}{k}=-i\eta \text{ \ (}\eta >0\text{)}
\end{equation*}%
thus $N$ is

\begin{equation*}
N=i\sqrt{\left( \frac{\eta ^{2}+\mathrm{\;sen}\;^{2}\theta _{1}}{1-\mathrm{%
\;sen}\;^{2}\theta _{1}}\right) }
\end{equation*}%
or we can define $N$ as

\begin{equation*}
N=i\beta
\end{equation*}%
where

\bigskip

\begin{equation*}
\beta =\sqrt{\left( \frac{\eta ^{2}+\mathrm{\; sen}\; ^{2}\theta _{1}}{1-%
\mathrm{\; sen}\; ^{2}\theta _{1}}\right) }
\end{equation*}

Substituting into Eq.(\ref{step})

\bigskip

\begin{equation*}
T=\left[ \frac{\frac{4i\beta }{\left( i\beta +1\right) ^{2}}}{1-\left( \frac{%
i\beta -1}{i\beta +1}\right) ^{2}e^{-2Ka}}\right] e^{-Ka-ik_{1}a}
\end{equation*}%
where 
\begin{equation*}
K=\frac{\sqrt{2m(V-E)}}{\hslash }
\end{equation*}

\bigskip For a thick barrier $ka>>1$ implies despise $e^{-2ka}$ in the
denominator, then:

\bigskip

\begin{equation*}
\Im =\left\vert T\right\vert ^{2}=\frac{16\beta ^{2}}{\left( \beta
^{2}+1\right) ^{2}}e^{-2ka}
\end{equation*}

As

\begin{eqnarray*}
\eta &=&\frac{k_{1}}{K}=\frac{\frac{i}{\hslash }\sqrt{2mE}}{\frac{i}{\hslash 
}\sqrt{2m(V-E)}} \\
\eta &=&\sqrt{\frac{E}{(V-E)}}\rightarrow \eta ^{2}=\frac{E}{(V-E)}
\end{eqnarray*}

\bigskip and

\begin{eqnarray*}
\beta ^{2} &=&\frac{E(1-\mathrm{\; sen}\; ^{2}\theta )V\mathrm{\; sen}\;
^{2}\theta }{(V-E)(1-\mathrm{\; sen}\; ^{2}\theta _{1})} \\
\beta ^{2}+1 &=&\frac{V}{(V-E)(1-\mathrm{\; sen}\; ^{2}\theta _{1})}
\end{eqnarray*}

\bigskip Finally we obtain

\begin{equation}
\Im _{\text{angular}}=\Im -\Im \mathrm{\;sen}\;^{2}\theta _{1}+\frac{8}{V^{2}%
}\left( V-E\right) ^{2}\mathrm{\;sen}\;^{2}(2\theta _{1})e^{-2ka}
\label{angular}
\end{equation}

\bigskip

Eq.(\ref{angular}) indicates two terms more on the equation and tunneling $%
\Im $ where we observe the sine function present. Thus, for $\theta =0^{0}$
we obtain the usual tunneling. The following graphs compare the usual
tunneling with the angular tunneling. The x-axis represents the particule
energy, whereas y-axis the tunneling probability. As we can see, there is a
range (between 30$^{\circ }$and 45$^{\circ }$) where the angular tunneling
is more favorable than the usual tunneling for energy below to 6 eV (Fig.
2). For $\theta =0^{0}$ we obtain the usual tunneling, i.e., the angular
tunneling is equal to the usual tunneling. For $\theta =90^{0}$ there is no
angular tunneling since the particle finds no obstacle Fig.2. It is
important to highlight that our results were obtained for a electron ranging
from 1 eV to 12 eV, striking a potencial barrier with 12 eV and 0.18 nm wide.

\section{Conclusion}

In this paper we have demonstrated that the incident angle influences the
probability of tunneling. According to our results there is a range (between
30$^{\circ }$and 45$^{\circ }$) where the angular tunneling is more
favorable than the usual tunneling for energy below to 6 eV. For $\theta
=0^{0}$ we obtain the usual tunneling, i.e., the angular tunneling is equal
to the usual tunneling. For $\theta =90^{0}$ there is no angular tunneling
since the particle finds no obstacle.

\textbf{Acknowlwdgement:} JHS thanks IFT and FAPESB Propp-00220.1300.1088
for financial support, CL thanks CAPES for support.

\end{document}